\documentclass[prd,aps,twocolumn,twoside,floatfix,showpacs]{revtex4}

\usepackage{epsfig}
\usepackage{amssymb}

\newcommand{\eg}{, {\it e.g.},\ }
\newcommand{\ie}{, {\it i.e.},\ }
\newcommand{\beq}{\begin{equation}}
\newcommand{\eeq}{\end{equation}}
\newcommand{\bea}{\begin{eqnarray}}
\newcommand{\eea}{\end{eqnarray}}
\newcommand{\nnl}{\nonumber \\}
\newcommand{\vs}{$\text{vs. }$}
\newcommand\eqn[1]{(\ref{eq:#1})}
\newcommand\fig[1]{Fig.~\ref{fig:#1}}

\newcommand\figurewidth{0.9\linewidth}

\begin{document}

\title{The observational legacy of preon stars -- probing new physics beyond the LHC}

\author{F. Sandin and J. Hansson}
\affiliation{Department of Physics, Lule{\aa} University of Technology, SE-97187 Lule\aa , Sweden}
\email{Fredrik.Sandin@gmail.com}

\begin{abstract}
We discuss possible ways to observationally detect the superdense cosmic
objects composed of hypothetical sub-constituent fermions beneath the
quark/lepton level, recently proposed by us.
The characteristic mass and size of such objects depend on the
compositeness scale, and their huge density cannot arise within
a context of quarks and leptons alone.
Their eventual observation would therefore be a direct vindication of
physics beyond the standard model of particle physics, possibly far beyond
the reach of the Large Hadron Collider (LHC), in a relatively simple and
inexpensive manner.
If relic objects of this type exist, they can possibly be detected by
present and future x-ray observatories, high-frequency gravitational wave
detectors, and seismological detectors.
To have a realistic detection rate\ie to be observable, they must
necessarily constitute a significant fraction of cold dark matter.
\end{abstract}


\pacs{12.60.Rc, 04.40.Dg, 95.35.+d}

\maketitle


\section{Introduction}

It is often assumed that cold dark matter (CDM) is some ``exotic'' type
of weakly interacting elementary particles, primordial relics created
in the early universe, not yet detected in particle accelerator experiments.
This hypothesis works well in cosmology, but both astrophysical
observations, and discrepancies between simulations and observations
of galaxies suggest that such a picture may be oversimplified.
For example, simulated density profiles of CDM halos are too cuspy,
more dwarf galaxies should have been observed because the number of
halos is expected to be inversely proportional to the mass,
and hydrodynamic simulations produce galaxy disks that are too small,
with too low angular momenta \cite{Ostrikeretal:2003}.
Moreover, there is a close relation between the rotation curve shape
and luminosity distribution in spiral galaxies, indicating that CDM
couples to luminous matter \cite{Sancisi:2003xt}, and the core
density in spiral galaxies is roughly constant, scaling with
the size of the core \cite{Gentile:2004tb}, in conflict with
predictions from such models. Further information and more examples
can be found in \cite{Zhitnitsky:2006vt} and references therein.

Considering the complexity of galaxies and the overall success of the
traditional view \cite{Concordance}\ie that CDM is composed of stable
weakly interacting (massive) particles (WIMPs), there are no truly
compelling reasons to abandon it.
It is sensible, however, to also explore alternative possibilities.
In particular, since there are indications that CDM couples to baryons,
parsimony (``Occam's razor'')
suggests that it could be a novel state of ``ordinary'' matter,
which decoupled from the radiation in the early universe before the onset
of primordial nucleosynthesis.
Any structure created at such an early epoch would necessarily
have a low characteristic mass and could therefore have remained unnoticed.

The spirit of this idea is not new. Already in the 1980s it was suggested
that lumps of stable
quark matter, so-called quark nuggets, could have formed in the early
universe \cite{Witten:1984rs}.
Should they exist, such objects contribute to CDM and if they were produced
in abundance they
could explain some observations that are inconsistent with the traditional
view \cite{Zhitnitsky:2006vt}. No observation precludes
the possibility that such objects compose the bulk of CDM, provided that
the mass of the objects does not exceed $\sim 10^{23}$~kg
\cite{Metcalf:2006ms,Zhitnitsky:2006vt}.
This idea has a natural extension to particle scales beneath the quark/lepton
level. Sub-quark particles (hereafter called preons) are motivated in
part by the existence of three fermion generations, and other unexplained
relations in the standard model of particle physics (SM), which indicate
that quarks and leptons could well be composite. Detailed motivations can
be found in\eg \cite{Dugne:2002fq} and references therein.
If preons exist, stable compact objects (``preon stars'') with densities
at least ten orders of magnitude higher than in quark nuggets/stars
could exist \cite{Hansson:04,Sandin:04}. See also \cite{Horvath:2007tr}
and \cite{Knutsen:1991,Cole:99}.
While the microscopic motivation for such objects is still somewhat schematic,
and the possibility that they formed in the early universe uncertain, it is
by no means impossible \cite{Nishimura:87,Horvath:2007tr}. As the consequences
of their eventual existence are very interesting and far-reaching, an
investigation of their phenomenology seems well-motivated.
In the present paper, we briefly discuss some possibilities to observe
compact preon dark matter (CPDM)\ie relic preon stars/nuggets, and how
the quark compositeness scale may be linked to astrophysical data.
A different scenario where dark matter is related to preons has been
suggested in \cite{Burdyuzha:1999df}.


\section{Properties and formation}

In the mid 1960s it was shown that for solutions to the stellar structure
equations, whether Newtonian or relativistic, there is a change in stability
whenever the mass reaches an extremum as a function of the
central density~\cite{Harrison:65}.
The instability in-between white dwarfs and neutron stars, which spans several
orders of magnitude of central densities, is an example of this property.
Consequently, beyond the density of the maximum mass neutron (or quark/hybrid)
star, $\sim 10^{16}$~g/cm$^3$ \cite{Kettner:95}, configurations are unstable.
The order of magnitude for this limiting density is valid also for a hypothetical
third class of compact stars \cite{Gerlach:68,Glendenning:00,Schertler:00} and
for stars composed of exotic hadron/quark condensates.
The instability is therefore generally assumed not to end before the Planck scale,
if at all.
This assumption, however, is valid only in the context of the SM, where quarks
and leptons are elementary.
If there is at least one deeper layer of constituents, beneath the particles of
the SM,
a corresponding class of stable compact objects could exist \cite{Hansson:04,
Sandin:04,Horvath:2007tr}. The density of such objects cannot be explained within
the context of the SM. This ``window of opportunity'' to new physics is our main
motivation for investigating means
to observe them. In the following, we briefly describe the
relation between the compositeness scale and the properties of such objects.

The characteristic density, size, and mass of a compact object depend on
the strength of the interactions between the constituent particles, see\eg
\cite{Narain:2006kx}. Qualitatively, the relation between these quantities
can be obtained in a simple way. Under the assumption that the equation of
state of matter is everywhere causal it follows that the radius, $R$, of a
stable compact object must exceed $4/3$ of its Schwarzschild radius,
$R_S = 2GM/c^2$, where $M$ is the mass of the object (without the assumption
of causality the factor is not $4/3$ but $9/8$), a result that follows
from the
general relativistic stellar structure equations. Simplifying the density
to be constant within
the object, this leads to an order of magnitude estimate for the relation
between the density, $\rho$, and the mass/radius of the maximum mass configuration
\bea
	M &\sim& \frac{9c^3}{64}\sqrt{\frac{2}{\pi G^3\rho}}, \\
	R &\sim& \frac{3c}{8}\sqrt{\frac{2}{\pi G\rho}}.
\eea
For neutron stars with $\rho\sim 10^{15}$~g/cm$^3$, this estimate yields
$M\sim 3$~M$_\odot$ and $R\sim 10$~km, correct order of
magnitudes for neutron stars.
We assume that the SM is reliable at least up to densities above the
onset of
the heaviest quark (top), which is of the order $\sim 10^{27}$~g/cm$^3$
for a charge-neutral fermion gas of six massive quarks and three massive
leptons with an
MIT bag constant chosen around the traditional value, $B^{1/4}\sim 150$~MeV.
The large mass of the top has been assumed to be a consequence of weak binding
between preons, see\eg \cite{Pati:1984jf}.
The phase where preons in the top quark can become deconfined should then
have a characteristic density
\beq
	\rho \sim \frac{m_t}{4/3\pi(\hbar c/\Lambda)^3}
	\simeq 9.5\times 10^{27}\,\text{g/cm}^3\,\left(\frac{\Lambda}{\text{TeV}}\right)^3,
	\label{eq:preondensity}
\eeq
where $m_t$ is the mass of the top quark, $\hbar c/\Lambda$ its ``size'',
and $\Lambda$ is expected to be of the order of the binding force scale parameter\ie
$\Lambda$ gives the compositeness energy scale.
Inserting this estimate in the expressions for the mass and radius of
the maximum mass configuration we obtain
\bea
	M &\sim& \frac{3}{32}\sqrt{\frac{6\hbar^3 c^9}{G^3 \Lambda^3 m_t}}
	  \simeq 2\times 10^{24}\,\text{kg}\,
	  \left(\frac{\text{TeV}}{\Lambda}\right)^{3/2},
	\label{eq:mass} \\
	R &\sim& \frac{1}{4}\sqrt{\frac{6\hbar^3 c^5}{G \Lambda^3 m_t}}
	  \simeq 3\times 10^{-3}\,\text{m}\,
	  \left(\frac{\text{TeV}}{\Lambda}\right)^{3/2}.
	\label{eq:radius}
\eea
Other estimates provided in \cite{Hansson:04,Sandin:04,Horvath:2007tr}
yield slightly different but qualitatively similar results.

\begin{figure}[ht!]
\epsfig{file=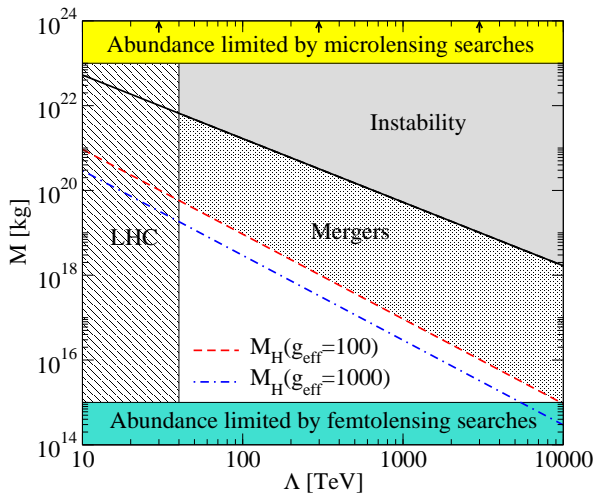,width=\figurewidth,clip=}
\caption{Constraints on the mass of compact preon dark matter (CPDM)
objects \vs the compositeness energy scale, $\Lambda$, which is related
to the length-scale of a composite top quark by $\hbar c/\Lambda$.
The estimate for the maximum mass of objects formed in the early universe,
$M_H(\Lambda,\,g_{\text{eff}})$, is the mass within the horizon at the
time of the preon phase transition, where $g_{\text{eff}}$ is the
effective number of degrees of freedom in the preon phase.
The LHC will probe compositeness scales up to about $40$~TeV.
The maximal mass of unobserved compact dark matter objects is $\sim 10^{23}$~kg
and femtolensing searches rule out $\sim 10^{14}<M<10^{15}$~kg.
No observational technique can presently resolve objects with
masses below $10^{14}$~kg. See the text for details.}
\label{fig:nuggets}
\end{figure}

CPDM objects could have been created in a first-order phase transition
in the early universe \cite{Sandin:04,Horvath:2007tr}, by a mechanism
similar to that described in \cite{Witten:1984rs}.
Under rather general assumptions, this scenario requires that the number
of microscopic degrees of freedom is higher during the preon era than
during the QCD/quark era \cite{Nishimura:87}.
This, perhaps counter-intuitive condition is satisfied by some preon
models and can be motivated by the simplicity of the representations
and the group structure, rather than an economic number of preons.
We do not further speculate about the details of the hypothetical phase
transition and the process of CPDM formation, as the main aim here is to
explore the possibility to detect such objects, if they exist.
We therefore assume that there was a first-order transition from a
preon phase to the quark/lepton phase, and that stable preon bubbles
formed. What would the characteristic mass of such bubbles be?
The density of the radiation background is
\beq
	\rho_R \simeq g_{\text{eff}}\,\frac{\pi^2}{30}\frac{(k_BT)^4}{\hbar^3c^5},
	\label{eq:raddensity}
\eeq
where $g_{\text{eff}}$ is the effective number of microscopic degrees
of freedom at temperature $T$.
Inserting \eqn{raddensity} in Friedmann's equations for a flat universe
(the curvature contribution anyway being negligible at early times) we
get an expression for the Hubble expansion parameter
\beq
	H \simeq \left[\frac{8\pi^3G}{90\hbar^3c^5}\,
	g_{\text{eff}}\right]^{1/2}(k_BT)^2.
\eeq
The maximum size of bubbles is limited by the event horizon\ie
the Hubble radius, $c/H$, at the temperature of the phase transition,
$T \simeq \Lambda/k_B$. The corresponding maximum mass of a preon
bubble is
\beq
	M_H \simeq \frac{4\pi}{3}\left(\frac{c}{H}\right)^3\!\!\rho_R
	\simeq 1.0\times 10^{24}\,\text{kg}\,g_{\text{eff}}^{-1/2}
	\left(\frac{\text{TeV}}{\Lambda}\right)^2,
	\label{eq:hubblemass}
\eeq
which is less than the maximum mass for stable objects \eqn{mass}.
From an observational point of view, $M_H$ is an estimate for the
maximal mass of CPDM objects, because the number of coalescence
events during the lifetime of the universe is negligible\eg from
\eqn{eventfraction}.
In reality, a typical preon bubble could be smaller or larger
than the Hubble radius at the critical temperature, depending
on the details of the phase transition and the bubble dynamics.
See \cite{Witten:1984rs} for a general discussion about formation
and evolution of quark bubbles in the QCD phase transition, and
\cite{Horvath:2007tr} for an analogous discussion about preon bubbles.
See also \cite{Hogan:1984hx} and \cite{Cottingham:1994ax}, where
different scenarios are discussed, leading to maximum masses of
quark bubbles that are, respectively, significantly smaller and
larger than the Hubble radius at the critical temperature.
For example, in \cite{Hogan:1984hx} it is suggested that the bubbles
should be smaller than the Hubble radius by a factor of at least
$\ln[(\hbar c^5/G)^{1/2}/(k_BT_{\text{QCD}})]/4$, which is about one
order of magnitude for $T_{\text{QCD}}\sim 150$~MeV.
A more precise estimate for the maximum mass of CPDM objects would
require further assumptions about the nature of preons and their
interactions, which are beyond scope of the present paper.
In \fig{nuggets} the estimates for the theoretical maximum mass
\eqn{mass} and the Hubble mass, $M_H$ \eqn{hubblemass}, are plotted \vs
the compositeness scale, $\Lambda$.
Included in the plot are also
the constraints on the mass of compact CDM from gravitational
lensing searches, see the next section. The Large Hadron Collider
(LHC) should allow exploration of compositeness scales up to
about $\Lambda\sim 40$~TeV, see \cite{Gianotti:2005ak}, where
future luminosity upgrades of LHC are discussed also.


\section{Gravitational lensing}

Gravitational lensing is today a well established field of astronomy, with 
a variety of astrophysical and cosmological applications. Among the many
interesting lensing phenomena, there is a possibility to observe low-mass
lenses by measuring interference effects between lensed images of
narrow astrophysical sources.
For lenses with masses in the range $\sim 10^{14}$~kg$\;<M<10^{17}$~kg,
the time delay induced by the lens would be comparable to the oscillation
period of
a gamma-ray. It has therefore been suggested \cite{Gould:92} that lenses
with masses in this range could be observed by gravitational lensing
of gamma-ray bursts (GRBs). Because the separation of the images would
be in the femto-arcsecond range for lenses and sources at cosmological
distances, this phenomenon is called ``femtolensing''. Femtolenses
would produce a characteristic pattern in the spectrum of GRBs
\cite{Stanek:93}, which is stable on time scales of 1~s, but might
slowly drift on time scales of 10~s due to the relative motion of the
lens and source.

No evidence for the existence of femtolenses presently exist,
but a number of GRB spectra, see \cite{Fenimore:88} and references
therein, have significant features that yet remain to be explained and
are similar \cite{GammaDetails} to those in a femtolensing model spectrum.
In particular, the GRB detector aboard the Ginga spacecraft
recorded ``absorption'' features with credible significance near 20 and
40~keV, especially for the burst GRB 880205 \cite{Murakami:88} and somewhat
less convincingly in the burst GRB 870303 \cite{Freeman:99}.
These features were originally interpreted as evidence for cyclotron
scattering of electrons in a strong magnetic field and, as a consequence,
a galactic origin of some GRBs, see\eg \cite{Fenimore:88,Wang:89}.
More recent observations (afterglows, supernova-GRB connection, etc.) and
theoretical models of GRBs falsify this explanation, in particular because
these were long GRBs, known to occur at cosmological distances.
The origin of the features observed with Ginga is therefore an
unsolved mystery. Similar features in the spectra of GRBs have
been detected in a number of other missions, notably at 11 and 35~keV
in GRB 890306 by Lilas \cite{Barat:93}, and at 50 and 70~keV in the
two peaks of GRB 780325 by HEAO A-4 \cite{Hueter:87}. Similar features have
been detected also by the BATSE spectroscopy detectors, see
\cite{Briggs:1999px} and references therein.

For more massive lenses, the energy-dependent spectra from a single
GRB detector provide no useful information. Instead, the spatial
interference effect needs to be measured. Two spacecrafts separated
by a distance that exceeds the radius of the Einstein ring of the lens,
$R_E\sim\sqrt{GM/(Hc)}\sim 10^7$m~$\times\sqrt{M/(10^{15}\;\text{kg})}$,
could detect lenses with masses in the range
$\sim 10^{15}$~kg$\;<M<10^{23}$~kg \cite{Nemiroff:1995ak}.
No present result limits the amount of CPDM with masses in this
range \cite{Marani:99}.
Consequently, refined femto- and picolensing searches could be used to
detect CPDM with masses in the range $10^{14}$~kg$\;<M<10^{23}$~kg.
A large abundance of CPDM with $M>10^{23}$~kg is, however, not consistent
with observations \cite{Metcalf:2006ms}. This does not preclude the
possibility that a small fraction of CDM is in that form, but since
the corresponding compositeness scale is within reach of the LHC,
see \fig{nuggets}, there is no reason to discuss that possibility here.
In the following, we briefly discuss the femtolensing effect on
the spectrum of GRBs.

\begin{figure}[ht!]
\epsfig{file=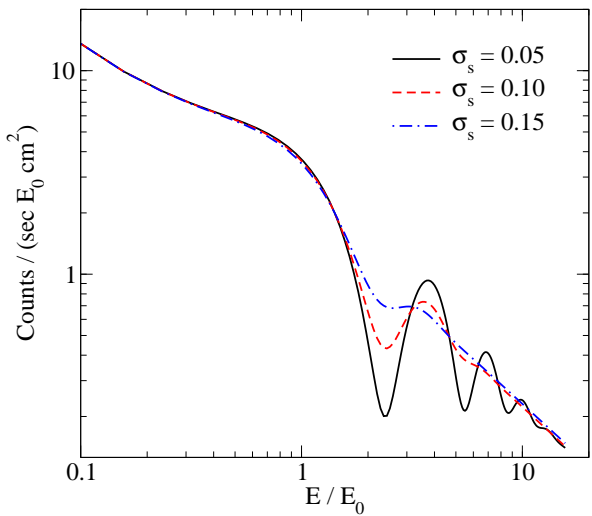,width=\figurewidth,clip=}
\caption{Femtolensing of a gamma-ray burst (GRB) with model spectrum
$(E/E_0)^{-1}$, for three different widths of the source, $\sigma_s$.
The GRB has a fixed
position relative to the optical axis, $r_s=0.5$, see text.
These spectra were calculated with the on-line interface
\cite{Sandin:femto}.}
\label{fig:femtolensing}
\end{figure}

\begin{figure}[ht!]
\epsfig{file=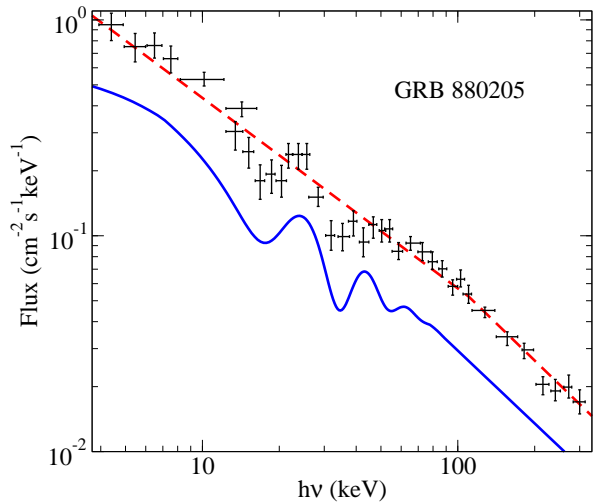,width=\figurewidth,clip=}
\caption{
Ginga spectral data of GRB 880205 for a power law model of the
incoming spectrum (dashed line), which is ruled out at more than $99.99$\%
confidence level~\cite{Fenimore:88}. The observed spectrum has line
features at $h \nu \simeq 20$ and $40$~keV, which could be due to
gravitational lensing (diffraction) of a Gaussian source by a
$\sim 10^{16}$~kg object at redshift $z \sim 1$ (solid line).
The spectral data depend on the model used and should not be directly
compared to the diffraction spectrum, which therefore has been shifted
downwards to enhance viewing \cite{GammaDetails}. The main concern here
is the location of the line features.}
\label{fig:grb880205}
\end{figure}

The magnification functions for point and extended sources have been
derived in \cite{Stanek:93}. These functions are not trivial to obtain
and have to be calculated numerically.
We have therefore provided an on-line tool \cite{Sandin:femto} for
calculation of femtolensing magnification functions and model GRB
spectra, which implements the model in \cite{Stanek:93} with some
extensions. The magnification function depends on four parameters,
the mass and redshift of the lens, the angular separation of the
source and lens, and the angular width of the source.
The width of the lens is neglected, because it has practically no
effect as long as the lenses are smaller than their Einstein ring.
We denote the angular diameter distances of the lens and the source
from the observer, and of the source from the lens with $d_L$, $d_S$,
and $d_{LS}$, respectively.
The distance, $r_s$, between the source and the optical axis is
measured in the dimensionless quantity
\beq
	r_s = \frac{\sqrt{\xi^2+\eta^2}}{d_s\theta_E},
\eeq
where $\theta_E=\sqrt{4GMd_{LS}/(c^2d_Ld_S)}$ is the angular radius
of the Einstein ring and $(\xi,\,\eta)$ are the Cartesian coordinates
of the source in the source plane. The dimensionless width of the source,
$\sigma_s$, is defined analogous to $r_s$\ie the actual width is divided
by $d_s\theta_E$.
Some femtolensing spectra are plotted in \fig{femtolensing},
for three different widths of a GRB, which is assumed to have
a fixed position relative to the optical axis, $r_s = 0.5$.
The model spectrum of the GRB is a assumed to be a power law, with
an exponent of $-1$. The energy scale depends on the redshift, $z$,
and mass, $M$, of the lens according to
\bea
	E_0 &=& \frac{hc^3}{4\pi G M(1+z)} \nnl
	&\simeq& {1.3\times 10^3\,\text{keV}}\,
	\left(\frac{10^{14}\,\text{kg}}{M}\right)
	\left(\frac{1}{1+z}\right),
\eea
for any model spectrum of the GRB.
In \fig{grb880205} a femtolensing spectra is superimposed on the 
spectral data of GRB 880205 for power law models of the GRB spectrum.

Because the amplitude of the femtolensing magnification function
decays with frequency (and the width of the source), detectors that
have energy thresholds well below the first minima should be used
in femtolensing searches.
According to Eq. (15) in \cite{Stanek:93}, the first minimum of the
magnification function is located at
\bea
	E_{1} &\simeq& \frac{3\pi E_0}{8r_s} \nnl
	&\simeq& {1.6\times 10^3\,\text{keV}}\,
	\left(\frac{10^{14}\,\text{kg}}{M}\right)
	\left(\frac{1}{1+z}\right)
	\left(\frac{1}{r_s}\right).
\eea
Spectra from the Transient Gamma-Ray Spectrometer and the BATSE
spectroscopy detectors used in recent searches for absorption line
features in GRB spectra were limited to $E>40$~keV and $20$~keV,
respectively, see \cite{Kurczynski:00,Briggs:1999px} and references
therein. Consequently, the advantages of these instruments fall
short in searches for femtolenses of high mass due to the relatively
high lower-energy thresholds.
The absorption features in GRB 870303, GRB 880205, and GRB 890306
observed in earlier missions would appear less significant if
observed with these instruments.
In particular, the low-energy absorption features in these bursts would
not be detected. The limit on the abundance of femtolenses given in
\cite{Marani:99} should therefore not be taken too seriously for more
massive femtolenses.
For masses in the picolensing range, there are
presently no limits on the abundance of CPDM (other than
$\rho_{\text{CPDM}}\leq\rho_{\text{CDM}}$).
A refined search for femto- and picolensing features in high-resolution
spectra of GRBs would therefore provide useful constraints on the
abundance of CPDM and similar compact dark matter objects.


\section{Gravitational waves from binaries}

While gravitational pico- and femtolensing can be used to detect and
estimate the mass spectrum of CPDM, these methods provide little information
about the actual density of the lenses ($R\lesssim d_L\theta_E$).
Consequently, lensing methods alone cannot provide detailed information
about the nature of the objects and their constituents.
One possibility to constrain the upper limit size of CPDM is
to measure high-frequency gravitational wave (GW) radiation
emitted from binary systems. In the following, we estimate the
properties and expected rate of such events for objects with
masses in the range $10^{15}$~kg$\;<M<10^{23}$~kg, which roughly is
the range unconstrained by gravitational lensing searches.

Assuming that the objects are distributed randomly in the solar
neighbourhood, the probability distribution function for the
semi-major axis, $a$, of binaries is \cite{Ioka:98}
\beq
	P(a)da = \frac{3}{4}\left(\frac{a}{\bar{x}}\right)^{3/4}
	\exp\left[-\left(\frac{a}{\bar{x}}\right)^{3/4}\right]\frac{da}{a},
	\label{eq:distribution}
\eeq
where $\bar{x}$ is the mean separation. Typically, the tidal
forces from nearby objects add angular momentum to a binary
and head-on collisions are thereby avoided.
We assume that the dark halo
density in the solar neighbourhood is 0.0079M$_\odot$~pc$^{-3}$
\cite{Alcock:2000ph}. For simplicity, we also assume that the bulk of
the dark halo is in the form of CPDM of equal masses. The results
can readily be generalised to an arbitrary fraction of CPDM.
The mean separation is
\bea
	\bar{x} &\simeq& \left(\frac{0.0079\text{M}_\odot}{M}
	\right)^{-1/3} \text{pc}.
	\label{eq:separation}
\eea
The remaining time before coalescence, $\tau$, due to emission of
GWs depends on the masses, the semi-major axis, and
the eccentricity of the orbit. For small $\tau$ the eccentricity can
be neglected, as the radiation reaction acts to reduce it. For
a circular orbit, the coalescence time is \cite{Gravitation}
\beq
	\tau = \frac{5 c^5}{512 G^3}\frac{a^4}{M^3}.
	\label{eq:tcoalescence}
\eeq
The probability distribution function \eqn{distribution} can be
expressed in the coalescence time $\tau$. Consequently, the relative
number of coalescence events within a time $t$ is obtained by
\bea
	\int_0^t P(\tau)d\tau
	&=& 1-\exp\left[
	-\bar{x}^{-3/4}\left(\kappa t \right)^{3/16}\right],
	\label{eq:eventfraction} \\
	\kappa &=& \frac{512 G^3 M^3}{5 c^5},
\eea
where $\int_0^\infty P(\tau)d\tau=1$.
The exponent in \eqn{eventfraction} is small for all masses
considered here, at any relevant timescale, $t$.
We therefore make the approximation $1-\exp(-x)\simeq x$.
The total number of objects, $N(D)$, within a distance $D$ can
be expressed in the local dark halo density and the mass of the
objects. The number of coalescence events, $N_c$, within a time
$t$ is $N_c = N(D)\int_0^tP(\tau)d\tau$, which yields
\beq
	N_c \simeq
	4.9 \left(\frac{D}{\text{pc}}\right)^3
	\left(\frac{10^{15}\,\text{kg}}{M}\right)^{11/16}
	\left(\frac{t}{\text{yrs}}\right)^{3/16}.
	\label{eq:eventrate}
\eeq
This estimate for the coalescence rate scales linearly with the
fraction of CPDM\ie there is an extra factor $\rho_{\text{CPDM}}/\rho_{\text{CDM}}$
on the right-hand side of \eqn{eventrate}.
In order to obtain a realistic event rate, a detector sensitive
enough to detect CPDM coalescence events at a distance of several
pc is needed. Next, we estimate the frequency and amplitude
of such events.

\begin{figure}[ht!]
\epsfig{file=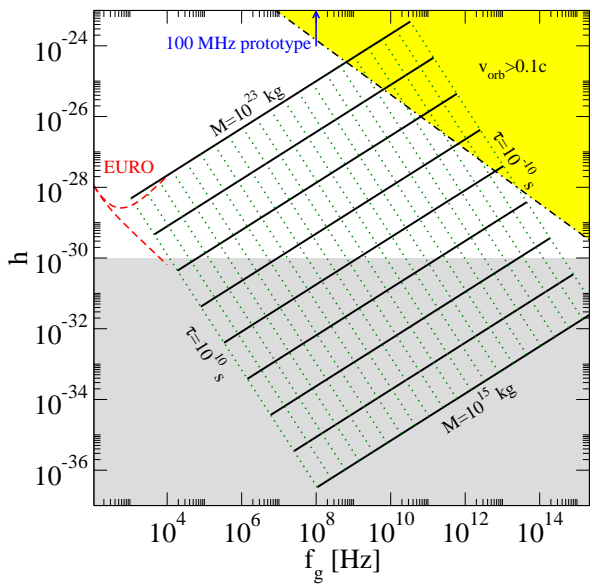,width=\figurewidth,clip=}
\caption{Amplitude \vs frequency for gravitational waves emitted
from an equal-mass binary system in circular orbit. The distance is
such that $10$ coalescence events per year is expected within that
range\ie $N_c=10$ in \eqn{eventrate}.
The solid lines denote the frequency-amplitude relation for different
masses, $M$, of the CPDM objects, in steps of one order of magnitude.
The coalescence time \eqn{tcoalescence} is denoted by the dotted
lines, also in steps of one order of magnitude.
The dash-dotted line corresponds to an orbital velocity of 10\% of
the speed of light.
Dashed lines denote the lower sensitivity curves for an observational time
of 5 years with EURO, according to two different design specifications
(shot-noise limited antenna with a knee-frequency of 1000 Hz and 
a xylophone-type interferometer).
The sensitivity of the first prototype 100 MHz detector in the UK
\cite{Cruise:2006zt} is presently insufficient to detect CPDM
coalescence events. The shaded region, $h<10^{-30}$, apparently is
beyond reach
of experiments and could be polluted by the relic gravitational
wave background, see \cite{Bisnovatyi-Kogan:2004bk} and references
therein.
}
\label{fig:gw}
\end{figure}

The frequency of GWs, $f_g$, emitted from a binary in a circular
orbit is twice the Kepler frequency
\bea
	f_g&=&\frac{1}{\pi}\left(\frac{2MG}{a^3}\right)^{1/2} \nnl
	&\simeq& 6.0\times10^{11}\,\text{Hz}\,\left(\frac{\text{sec}}
	{\tau}\right)^{3/8}\left(\frac{10^{15}\,\text{kg}}{M}\right)^{5/8},
\eea
because the waves are essentially generated by the quadrupole
moment of the binary. The power emitted in GWs is \cite{Gravitation}
\bea
	L_g&=&\frac{64G^4}{5c^5}\left(\frac{M}{a}\right)^5 \nnl
	&\simeq& 1.4\times 10^{16}\,\text{W}\,
	\left(\frac{M}{10^{15}\,\text{kg}}
	\frac{f_g}{\text{GHz}}\right)^{10/3},
\eea
and the amplitude of the GWs at a distance $D$ from the source is
\bea
	h &=& \left(\frac{G L_g}{\pi^2 c^3}\right)^{1/2}\frac{1}{f_g D} \nnl
	&\simeq& 1.9\times10^{-36}
	\left(\frac{M}{10^{15}\,\text{kg}}\right)^{5/3}
	\left(\frac{f_g}{\text{GHz}}\right)^{2/3}
	\left(\frac{\text{pc}}{D}\right).
	\label{eq:gwamplitude}
\eea
The frequency dependent amplitude \eqn{gwamplitude} is plotted in
\fig{gw} for different masses, $M$, and distances, $D$, chosen such
that 10 coalescence events per year are expected with at least that
amplitude (if the CPDM fraction of CDM is less than one, the number of
events per year is lowered by the same factor).
Also indicated in the plot are the coalescence time \eqn{tcoalescence},
the threshold of the relativistic domain, where the orbital velocity,
$v_{orb}=c(R_S/a)^{1/2}$, is 10\% of the speed of light, and an estimate
for the sensitivity of future detectors, $h_{\text{min}}\sim 10^{-30}$,
see \cite{Bisnovatyi-Kogan:2004bk} and references therein.
Even if this estimate for the sensitivity could be exceeded, 
coalescence events with significantly lower amplitudes would be
difficult to distinguish from the stochastic GW background,
created by quantum fluctuations in the early universe. This background
exists in most popular cosmological models and, due to the expansion of
the universe, the amplitudes of the initial fluctuations are amplified
and should approach the $h\sim 10^{-30}$ level \cite{Bisnovatyi-Kogan:2004bk}.
Because $h\propto N_c^{-1/3}$, the amplitudes in \fig{gw} will increase
only by a factor two for an order of magnitude decrease of the
event rate. We therefore choose $N_c=10$, to compensate for the
simplifying assumption that $\rho_{\text{CPDM}}/\rho_{\text{CDM}}=1$.

The planned spectral noise density for the European Gravitational Wave
Observatory (EURO) in the range 10-10000~Hz is \cite{Euro}
\bea
	S_n(f) &=& 10^{-50} \Bigg[
	\left(\frac{f}{245\,\text{Hz}}\right)^{-4}+
	\left(\frac{f}{360\,\text{Hz}}\right)^{-2} \nnl
	&&+\left(\frac{f_k}{770\,\text{Hz}}\right)
	\left(1+\frac{f^2}{f_k^2}\right)\Bigg]
	\,\text{Hz}^{-1},
	\label{eq:euro}
\eea
where $f_k=1000$~Hz is the knee frequency. Alternatively, EURO will be
based on a xylophone-type interferometer, which has higher sensitivity
at high frequencies. The spectral noise density for the latter choice
is described by \eqn{euro} when the last $f_k$-dependent term is omitted.
The characteristic amplitude of a GW is $h_c=h\sqrt{n}$, where
$n=f_g\Delta T$ is the number of cycles during an observational time of
$\Delta T$. The wave strength of GWs from a monochromatic source observed
with an interferometer is $h_s = h_c/\sqrt{5f_g}$. Consequently, the
minimum amplitude, $h_{\text{min}}$, that can be resolved with EURO after
an observational time $\Delta T$ is $h_{\text{min}}=\sqrt{5S_n(f)/\Delta T}$.
This estimate for the sensitivity of EURO is plotted in \fig{gw} for
an observational time of five years.
High-mass CPDM is marginally within range of the next generation of
gravitational wave detectors. However, the semi-major axis of a binary is
\bea
	a &=& R_S\left(\frac{c^3}{2\pi GMf_g}\right)^{2/3} \nnl
	&\simeq& 1.6\times 10^7 R_S\left(
	\frac{10^{15}\,\text{kg}}{M}
	\frac{\text{GHz}}{f_g}\right)^{2/3},
\eea
so in order to get useful constraints on the compactness of CPDM, a
detector sensitive at higher frequencies is needed.
Interestingly, high-frequency GW detectors are laboratory-scale
devices that are relatively inexpensive to construct. A first 100~MHz
prototype has recently been built in the UK \cite{Cruise:2006zt}.
If the sensitivity of such detectors would approach the estimates
given in \cite{Bisnovatyi-Kogan:2004bk}, they would provide useful
constraints on CPDM.
The range $10^{15}$~kg$\;<M\lesssim10^{18}$~kg would, however, only
be accessible by rare nearby events.


\section{Conclusion \& Discussion}

If quarks and leptons are composite particles, superdense preon stars (or
``nuggets'') could exist \cite{Hansson:04,Sandin:04,Horvath:2007tr}.
While the microscopic motivation for such objects is still somewhat schematic
and the exact process of formation uncertain, the consequences of their eventual
existence are far-reaching.
In the present paper we briefly investigate their phenomenology, assuming
that they formed in the early universe and contribute significantly to CDM.
Their maximum mass is roughly limited by the horizon at the time
of formation, $M_H\sim 10^{24}\,\text{kg}\,g_{\text{eff}}^{-1/2}(\text{TeV}/\Lambda)^2$,
where $\Lambda$ is the quark compositeness energy scale and $g_{\text{eff}}$
is the number of microscopic degrees of freedom in the primordial preon phase.
This is a factor $\sim 2\sqrt{g_{\text{eff}}\,\Lambda/\text{TeV}}$ lower than
the maximum mass for stable hydrostatic configurations. However, the typical
mass could be lower or higher than this estimate, depending on the properties
of preons and their interactions.
Gravitational lensing searches put strong constraints on the abundance
of CDM objects with masses in the ranges $M\gtrsim 10^{23}$~kg and
$10^{14}\lesssim M\lesssim 10^{15}$~kg.
Unexplained features in GRB spectra observed by\eg Ginga, Lilas, and
HEAO A-4 motivate a continued search for gravitational pico- and femtolenses.
This would provide useful constraints on the abundance of compact CDM objects with
masses in the range $10^{15}\lesssim M\lesssim 10^{23}$~kg, corresponding
to a maximum compositeness energy scale for CPDM of a few thousand TeV.
This observational technique, however, provides little information about
the nature of the lenses, because their size and density is limited only
by the radius of their Einstein ring.
Future high-frequency gravitational wave detectors could provide complementary
information
about the density of compact CPDM binaries, but it is presently unclear whether
it is possible to detect the chirp signal of a low-mass binary as the objects
coalesce \cite{MikeCruise}. This would be necessary in order to obtain a useful
constraint on the radii and, consequently, a lower-limit for the density of
the objects.
In an optimistic scenario, where the mass of the CPDM objects is comparable to
$M_H$, this method could be useful to indirectly detect compositeness
up to a few hundred TeV.
Another possibility to detect compact CDM objects and to constrain their
density is by seismology\ie by measuring the seismic waves generated as they
pass through the Earth or the Moon, see \cite{Herrin:2005kb} and references
therein.
Unlike the observational methods discussed above, this method is useful in
scenarios where the typical mass of the objects is low, as the collision
rate increases with the number density of objects. The cross-section of a
CPDM object would be at least six orders of magnitude smaller than for a
quark nugget of equal mass, making it possible to distinguish them.
Should CPDM objects exist, their observational detection may well be the only
means, for quite a long time, to discover compositeness beyond the reach
of the LHC and other near-future accelerators. As the observational techniques
discussed here are useful also for other purposes, and are already in operation
to some extent, they constitute a comparatively simple and inexpensive way to
test the CPDM hypothesis.


\acknowledgments{
F.S. acknowledges support from the Swedish Graduate School of
Space Technology and J.H. acknowledges support from Carl Tryggers
stiftelse.}



\begin{thebibliography}{00}
\bibitem{GammaDetails}
The observed counts in a gamma-ray detector corresponds to the energy
loss spectrum, which is created by the gamma photons as they interact
with the medium in the detector\eg sodium iodide or germanium.
Consequently, it is not certain that the energy of a particular
count equals the energy of an incident photon. The spectrum of the
light entering the detector can therefore not be deduced from data.
Instead, model spectra are used in conjunction with data to calculate
the probability that the physical spectrum is/is not explained by a
particular model. In this procedure, details about the detector
response functions and the original data are needed.
We were unable to obtain this information for the detector aboard
Ginga and for GRB 880205. A formal detailed comparision with femtolensing
spectra has therefore not been done.
\end{thebibliography}
\end{document}